\documentclass[aps,preprint,showpacs]{revtex4}
\usepackage{graphicx}
\usepackage{amsmath}
\newcommand{\be}{\begin{equation}}
\newcommand{\ee}{\end{equation}}
\begin{document}

\title{Surface solitons in two-dimensional chirped photonic lattices}

\author{Mario I. Molina$^{1,2,3}$,  Yaroslav V. Kartashov$^2$, Lluis Torner$^2$, and
Yuri S. Kivshar$^3$}

\affiliation{$^1$ Departamento de F\'{\i}sica, Facultad de Ciencias,
Universidad de Chile, Santiago, Chile\\ $^2$ICFO-Institut de
Ciencies Fotoniques, and Universitat Politecnica de Catalunya,
Mediterranean Technology Park, 08860 Castelldefels (Barcelona), Spain\\
$^3$Nonlinear Physics Center, Research School of Physical Sciences
and Engineering, Australian National University, Canberra, ACT 0200,
Australia}

%\date{\today}

\begin{abstract}
We study surface modes in semi-infinite chirped two-dimensional
photonic lattices in the framework of an effective discrete
nonlinear model. We demonstrate that the lattice chirp can change
dramatically the conditions for the mode localization near the
surface, and we find numerically the families of surface modes, in
linear lattices, and discrete surface solitons, in nonlinear
lattices. We demonstrate that, in a sharp contrast to
one-dimensional discrete surface solitons, in two-dimensional
lattices the mode threshold power is lowered by the action of both
the surface and lattice chirp. By manipulating with the lattice
chirp, we can control the mode position and its localization.
\end{abstract}


\maketitle

\section{Introduction}

In linear guided-wave optics surface states have been predicted to
exist at interfaces separating periodic and homogeneous dielectric
media~\cite{optics}. The interest in the study of surface states has
been renewed recently because an interplay of self-focusing
nonlinearity and repulsive effect of the surface was shown to
facilitate the formation of discrete surface solitons located at the
edge of the waveguide array~\cite{makris,suntsov}. These surface
solitons can be understood as discrete optical solitons localized
near the surface provided their power exceeds a certain threshold
value, compensating the repulsive action of the
surface~\cite{surfol}. A similar effect of the light localization
near the edge of the waveguide array and the formation of surface
gap solitons have been predicted and observed for defocusing
nonlinear periodic
media~\cite{makris,PRL_kartashov,PRL_canberra,kip}.

In recent studies~\cite{chirp_bar}, it was demonstrated that the
conditions for the soliton formation can be dramatically modified
near the surface of a chirped optical lattice. It was found that,
due to combined actions of internal reflection at the interface,
distributed Bragg-type reflection and focusing nonlinearity,
surfaces of chirped lattices become soliton attractors, in a sharp
contrast with the standard lattices. The main conclusions of those
studies have been confirmed later by employing a discrete
model~\cite{chirp_our}.

It is important to analyze how the properties of nonlinear surface
waves are modified by the lattice dimensionality, and the first
studies of different types of discrete surface solitons in
two-dimensional nonlinear photonic
lattices~\cite{ol_kar1,ol_kar2,ol_moti,pre_arxiv,pla_our} revealed,
in particular, that the presence of a surface increases the
stability region for two-dimensional (2D) discrete
solitons~\cite{pre_arxiv} and the threshold power for the edge
surface state is slightly higher than that for the corner
soliton~\cite{ol_moti,pla_our}. Recent observations of
two-dimensional surface solitons in optically-induced photonic
lattices~\cite{2d_exp_1} and laser-written waveguide arrays in fused
silica~\cite{2d_exp_2} demonstrated novel features of these
nonlinear surface modes in comparison with their counterparts in
one-dimensional waveguide arrays.

In this paper, we employ a two-dimensional discrete model of a
chirped photonic lattice to provide a deeper physical insight into
the properties of linear surface modes and discrete surface solitons
in two-dimensional chirped photonic lattices. In particular, we
demonstrate that there appears a critical value of the lattice chirp
for the existence of a linear mode localized at the surface such
that above this critical value the surface solitons originate from
the corresponding linear surface modes, and they do not require any
threshold power for their existence. We study the dependence of the
mode position near the surface and the critical power as a function
of the lattice chirp, and also demonstrate how the engineered chirp
of the photonic lattice can facilitate a selective nonlinear
localization near the surface.

The paper is organized as follows. In Sec. II we introduce our
discrete nonlinear model of a two-dimensional chirped lattice.
Section III is devoted to the study of localized modes in linear
chirped lattices. In particular, we consider two special cases of
the chirp and demonstrate that the chirp can change dramatically the
conditions for the mode localization near the surface, so that
surface modes may exist even in a linear lattice provided the chirp
parameter exceeds some threshold value. In Sec. IV we study the
localization of nonlinear modes near the surface and, in particular,
demonstrate that by engineering the lattice chirp we can control the
localization of nonlinear modes and their location on the lattice.
Finally, Sec.~V concludes the paper.

\section{Model}

We consider a two-dimensional array of $N\times N$ nonlinear (Kerr)
single-mode waveguides which form a square lattice with the lattice
spacing $a (\equiv 1)$. We introduce a spatial chirp into the system
by either changing the individual refraction index of the guides
and/or adjusting the waveguide spacings. In the first case, both the
propagation constant and coupling between the neighboring waveguides
become chirped; in the second case, only the coupling is affected.
In particular, it is possible to have a situation where the coupling
chirp is greatly reduced by compensating a change in the refraction
index by a judicious change of the waveguide distance. As a result,
only chirp in the propagation constant is relevant in this case. In
what follows we consider both the cases separately since they do
lead to a different asymptotic behavior.

In the framework of the coupled-mode theory, the nonlinear equations
for the amplitudes of the stationary modes  can be written in
general as
\be
-\beta C_{{\bf n}} + \lambda_{{\bf n}} C_{\bf n}+   \sum_{{\bf m}}
V_{{\bf n},{\bf m}}\ C_{\bf m} + \gamma |C_{\bf n}|^2 C_{\bf n} =
0\label{eq:1}
\ee

For a two-dimensional semi-infinite square nonlinear lattice,
Eq.(\ref{eq:1}) becomes
\begin{align}
 &-\beta C_{n,m} + \lambda_{n,m} C_{n,m} + \sum_{j=\pm 1}
V_{(n,m), (n+j,m)}C_{n+j,m} \nonumber \\
& +\sum_{j=\pm 1} V_{(n,m), (n,m+j)}C_{n,m+j} + \gamma |C_{n,m}|^2
C_{n,m}  =  0, \label{eq:2}
\end{align}
where $n\ge 0$ and $m\ge0$, and $V_{(n,m),(n',m')}=0$ if either
$n'<0$ or $m'<0$.

\begin{figure}[t]
\noindent\includegraphics[scale=0.37]{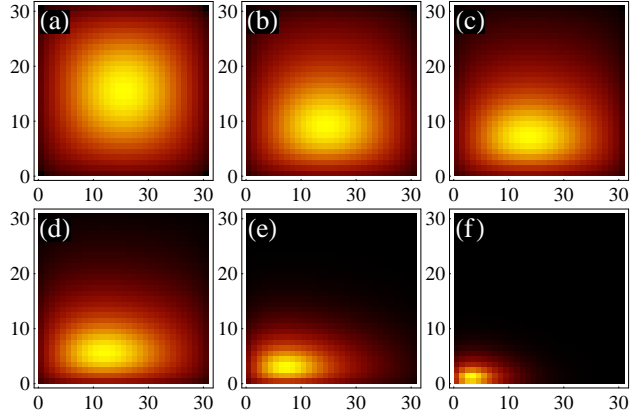} \caption{Density
profiles of localized nodeless modes for different values of the
chirp ($\sigma_{x}, \sigma_{y}$) for a $31\times 31$ linear array
with $\lambda_{0}/V_{0}=5$. (a): $(0,0)$, (b) $(5\times
10^{-5},5\times 10^{-4})$, (c) $(10^{-4},10^{-3})$, (d) $(2\times
10^{-4},2\times 10^{-3})$, (e) $(10^{-3},10^{-2})$, (f)
$(10^{-2},10^{-1}).$ Anisotropic chirp in the propagation constant.}
\label{figura1}
\end{figure}

We now modulate the refraction index $\rho(n,m)$ of each guide to
decrease it exponentially away from the corner site $(0,0)$, in a
general, anisotropic manner:
\[\rho(n,m) = \rho_{0} \exp(-\sigma_{x} n - \sigma_{y} m),\]
where $\sigma_{x}$ and $\sigma_{y}$ are two chirp parameters in the
$x$ and $y$-direction, respectively. This change in the refraction
index affects the effective propagation constant of each guide in a
similar way,
\be \lambda_{n,m} = \lambda_{0} \exp(-\sigma_{x} n - \sigma_{y}
m)\label{eq:3} \ee
In principle, the couplings between the neighboring waveguides is
also affected by the chirp. However, if the chirp in the refraction
index varies slowly in space, then in the first-order approximation
the waveguide couplings is affected in the same way:
\[V_{(n,m),(n+1,m)} = V_{0} \exp(-\sigma_{x} n - \sigma_{y} m) =
V_{(n,m),(n,m+1)}
\]
\[V_{(n,m),(n-1,m)} = V_{0} \exp(-\sigma_{x} (n-1) - \sigma_{y} m)
\]
\[V_{(n,m),(n,m-1)} = V_{0} \exp(-\sigma_{x} n - \sigma_{y} (m-1))
\]

However, since the coupling between the waveguides also depends upon
an overlap of the mode fields, it is possible in principle to alter
the waveguide spacing judiciously in order to compensate for the
spatial variation of the refraction index, rendering the couplings
nearly constant: $V_{{\bf n},{\bf m}}\approx V_{0}$. Hereafter, we
consider both the cases separately.

\begin{figure}[t]
\noindent\includegraphics[scale=0.37]{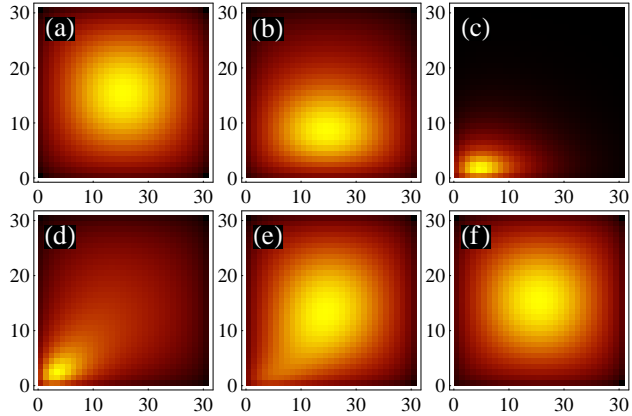} \caption{Density
profiles of localized nodeless mode for different values of the
chirp ($\sigma_{x}, \sigma_{y}$) for $\lambda_{0}/V_{0}=1$. (a):
$(0,0)$, (b) $(2\times 10^{-4},3\times 10^{-3})$, (c) $(3\times
10^{-2},3\times 10^{-1})$, (d) $(2\times 10^{-1},5\times 10^{-1})$,
(e) $(10^{-1},1)$, (f) $(1,1)$. Anisotropic chirp in the propagation
constant only. } \label{figura2}
\end{figure}

\section{Linear surface modes}

\subsection{Chirp in the propagation constant}

Equations for the stationary modes have the form,
\begin{align} &\beta C_{n,m} = \lambda_{0} \exp(-\sigma_{x} n -
\sigma_{y} m)\ C_{n,m} \nonumber \\
&+ V_{0} \sum_{j=\pm 1} (C_{n+j,m}+ C_{n, m+j}) + \gamma
|C_{n,m}|^2 C_{n,m},\label{eq:7}
\end{align}
We start by examining the $N\times N$ linear modes ($\gamma=0$),
relevant to the case of weak input power $\sum_{n,m}
|C_{n,m}|^{2}\ll 1$. For a fixed number of waveguides, we
diagonalize the system of discrete linear equations (\ref{eq:7}) and
find all the modes. In particular, we focus on the nodeless
(fundamental) mode, and how it changes as a function of the chirp
parameters $\sigma_{x}$ and $\sigma_{y}$. From Eq.~(\ref{eq:7}) we
find that, in the limit of large chirp, $\sigma_{x},\sigma_{y} \gg
1$, the system reduces to a two-dimensional lattice with a linear
impurity at one of the corner sites. It can be
proven~\cite{prb_molina} that in this case for $N\rightarrow \infty$
a localized mode centered at the corner site is possible, provided
\[ \lambda_{0}/V_{0} > [2-(16/3 \pi)]^{-1}\equiv
(\lambda_{0}/V_{0})_{c}. \]
Thus, there appear two regimes: (1) For $\lambda_{0}/V_{0}
> (\lambda_{0}/V_{0})_{c}\sim 3.3$, an increase in the chirp values
shifts the center of the nodeless mode from the center of the
lattice towards the corner site. For sufficiently large average
chirp, $<\sigma> = (1/2)(\sigma_{x} + \sigma_{y})$, the mode center
approaches the corner site position and its spatial extension
reduces considerably (see Fig.~1). (2) When $\lambda_{0}/V_{0} <
(\lambda_{0}/V_{0})_{c}$, the mode center shifts from the center of
the lattice towards the corner site upon the average chirp increase
as above, but upon further chirp increase, the mode center shifts
away from the corner site and it approaches the center of the
lattice asymptotically (see Fig.~2).

\begin{figure}[h]
\noindent\includegraphics[scale=0.52]{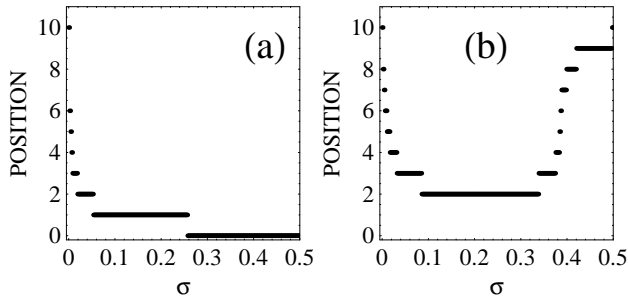} \caption{Distance
of the mode maximum from the corner of the array as a function of
the chirp parameter, for the isotropic case
$\sigma_{x}=\sigma_{y}\equiv \sigma$. (a): $\lambda_{0}/V=5$. (b):
$\lambda_{0}/V=1$.} \label{figura3}
\end{figure}

In order to visualize and quantify this behavior of the mode center
as the chirp is increased, we consider the isotropic case
$\sigma_{x} = \sigma_{y}$ and calculate numerically the mode
position relative to the lattice corner along the diagonal $(n,n)$
as a function of the chirp for both the cases, $\lambda_{0}/V_{0} >
(\lambda_{0}/V_{0})_{c}$ and $\lambda_{0}/V_{0} <
(\lambda_{0}/V_{0})_{c}$.  The results are shown in Fig.~3. The
observed dependence is presented by a series of discrete steps of a
widely varying width. At small chirp, the fundamental mode is
confined well inside the lattice and a small change of the chirp can
alter the position of the mode maximum considerably. As the mode
gets closer to the corner, the steps increase in size. For
$\lambda_{0}/V_{0} > (\lambda_{0}/V_{0})_{c}$, the mode finally
reaches the corner site where it will remain upon further chirp
increase see Fig.~3(a)]. On the contrary, for $\lambda_{0}/V_{0} <
(\lambda_{0}/V_{0})_{c}$, the mode reaches a minimum distance from
the corner and, upon further chirp increase, it shifts away from the
corner and reaches the middle of the lattice in the limit of very
large chirp [see Fig. 3(b)].

\begin{figure}[h]
\noindent\includegraphics[scale=0.37]{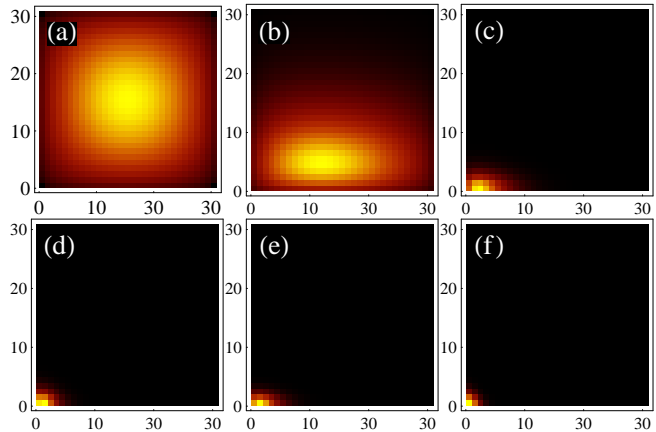} \caption{Same as
in Fig.2 but including chirp in propagation constant {\em and} in
waveguide couplings.} \label{figura4}
\end{figure}

\begin{figure}[h]
\noindent\includegraphics[scale=0.3]{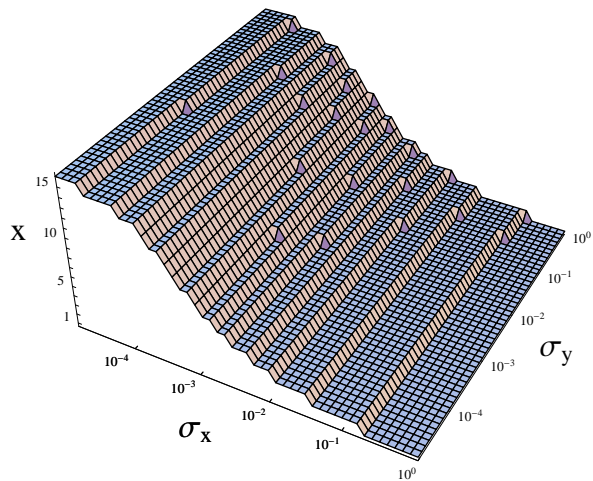} \caption{Position
of the center of the nodeless mode along the $x$-coordinate as a
function of the chirp parameter for $N=21$ and
$\lambda_{0}/V_{0}=1$. The case of an anisotropic chirp in both the
propagation constant and couplings. } \label{figura5}
\end{figure}

\subsection{Chirp in the propagation constant and waveguide coupling}

This seems to be more realistic case that originates from modulation
of the refractive index imposed on the periodic square array of
coupled waveguides. If the modulation is smooth, then both the
propagation constants and waveguide couplings are affected in the
same manner. This leads to the following set of coupled equations
for the linear modes:
\begin{align}
&\beta C_{n,m} = e^{-\sigma_{x}n-\sigma_{y} m} [\lambda_{0} C_{n,m}
+
V_{0} (C_{n+1,m} +C_{n,m+1})] \nonumber\\
&+ V_{0} e^{-\sigma_{x}n-\sigma_{y} m} [e^{\sigma_{y}} C_{n,m-1} +
e^{\sigma_{x}} C_{n-1,m}].
 \label{eq:ultima}
\end{align}
where $C_{n',m'}=0$ if either $n'<0$ or $m'<0$.

%%%%%%%%%%%%%%%%%%%%%%%%%%%%%%%%%%%%%%%%%%%%%%%%%%%%%%%%%%%%%

In this case, from Eq.~(\ref{eq:ultima}) we obtain that, in the
limit of a large average chirp, the system becomes an effective
trimer. Thus, the localized nodeless mode remains in the immediate
vicinity of the corner site, and there will not be any bouncing
phenomena for the mode center as in the previous section.

As the values of the chirp in both directions are increased from
zero, the fundamental mode will shift from the center of the lattice
approaching the corner site, as shown in Fig.~4. The precise route
of the mode center towards the corner site depends on the values of
both $\sigma_{x}$ and $\sigma_{y}$. We now perform a numerical sweep
in the $(\sigma_{x},\sigma_{y})$ space and record the position
$(x,y)$ of the mode maximum. The results for the $x$-position are
shown in Fig.~5

Due to symmetry, the position of the y-coordinate of the mode
maximum can be obtained from the same figure by exchanging
$\sigma_{x}$ and $\sigma_{y}$. We notice that, at small average
chirp value, the mode center is very sensitive to the precise value
of the chirps. As chirps are increased, the position of the mode
center becomes less sensitive and the width of the `terraces' become
larger. For chirp values greater than, say $\approx 0.3$, the mode
will be centered on the corner site and will remain there upon
further chirp increase.

\section{Surface solitons in chirped nonlinear lattices}

Having examined the main features of the fundamental linear (i.e.,
low power) surface modes in the presence of an anisotropic chirp, we
now consider the case of a nonlinear lattice, i.e. $\gamma \neq 0$.
We focus on the more interesting case examined in the previous
section and, for a given value of $\beta$, we find nonlinear
localized states by solving the stationary equations numerically
with the help of a straightforward extension of the multidimensional
Newton-Raphson method used earlier in our analysis of the
one-dimensional system~\cite{chirp_our}. To do so, we first present
the two-dimensional $N\times N$ system as an effective, $N^2$-long
one-dimensional chain by means of a convenient relabeling of the
site coordinates, and then apply the usual Newton-Raphson
method~\cite{chirp_our}.

As an example, we consider the linear modes presented in
Figs.~4(a-f) and recalculate them in the case of nonlinear lattice
for $\gamma=1$. The corresponding results for the nonlinear
localized modes are shown in Figs.~6(a-f). We note that, in
comparison with the linear modes, the position of the nonlinear mode
center does not change, but the spatial extension of the mode is
reduced considerably, due to the self-trapping effect.

\begin{figure}[h]
\noindent\includegraphics[scale=0.37]{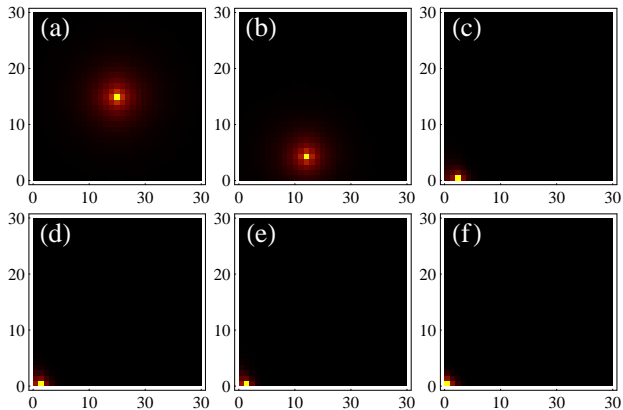}
\caption{Nonlinear localized modes near the corner of a
two-dimensional lattice in the case of an anisotropic chirp in both
the propagation constant and couplings. Parameters are the same as
in Fig.~4 but for $\gamma=1$.} \label{figura6}
\end{figure}

Finally, we calculate the minimum power needed to create a nonlinear
localized mode at the corner site, for given values of the chirp
parameters $(\sigma_{x},\sigma_{y})$. As we have might easily
conjecture, at small average values of the lattice chirp, the
minimum power is needed to create a stable mode while above a
certain value of the chirp, no minimum power is required, similar to
the case of a one-dimensional chirped lattice~\cite{chirp_our}.
Figure 7 shows that this is indeed the case.

\begin{figure}[h]
\noindent\includegraphics[scale=0.45]{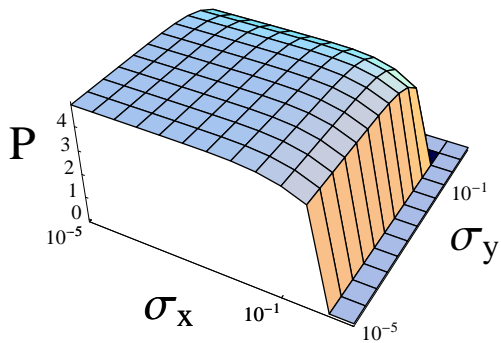} \caption{ Minimum
power for creating a nonlinear localized surface mode at the lattice
corner, as a function of the chirp parameters
$(\sigma_{x},\sigma_{y})$ for $\lambda_{0}/V=1$. The case of an
anisotropic chirp in both the propagation constant and couplings. }
\label{figura7}
\end{figure}

Therefore, by changing either the lattice chirp or the effective
nonlinearity, we can control not only the mode localization but also
its location in the lattice near the surface. This property provides
an efficient physical mechanism for the low-power generation of
nonlinear localized modes near lattice surfaces and other
defects, where the threshold power for the generation of nonlinear
localized mode is lower than in a bulk. It is now possible to move the localized
mode in a given direction up to a desired position on the lattice
by engineering both, the value and direction of the lattice chirp,
and then localize the mode by further increasing the nonlinearity.
As a matter of fact, this simple mechanism is presented in
Figs.~4(a-f) and Figs.~6(a-f).

\section{Conclusions}

We have analyzed the properties of surface modes and discrete
surface solitons in two-dimensional chirped linear and nonlinear
photonic lattices, in the framework of an effective two-dimensional
discrete model. We have demonstrated that the lattice chirp can
change dramatically the conditions for the mode localization near
the surface, and we have found numerically the families of linear
surface modes and discrete surface solitons. We have demonstrated
that, by manipulating the lattice chirp, we can control not only the
mode localization but also its location in the lattice near the
surface, and thus provide an efficient tool for the low-power
generation of nonlinear localized modes and their control in
realistic physical systems. We believe the basic principles of
engineering the localization of linear and nonlinear localized modes
in chirped lattices can be useful for their implementation in other
discrete nonlinear systems.

\section{Acknowledgements}

This work has been supported by Fondecyt grants 1050193 and 7050173
in Chile, and by the Australian Research Council in Australia. M.I.M.
thanks ICFO-Institut de Ciencies Fotoniques (Bercelona) and the
Nonlinear Physics Center at the Australian National
University for hospitality and support.

\end{document}